# *In vitro* evaluation of a novel Mg–Sn–Ge ternary alloy for orthopedic applications

Xian Wei[a,b,c], Sujie Ma[b], Jiajia Meng[b], Hong Qing[a], and Qing Zhao[b,d,1]

[a] School of Life Science, Beijing Institute of Technology, Beijing 100081, China

[b] Center for Quantum Technology Research and Key Laboratory of Advanced Optoelectronic Quantum Architecture and Measurements (MOE), School of Physics, Beijing Institute of Technology, Beijing 100081, China

[c] Department of science, Taiyuan Institute of Technology, Taiyuan 030008, China

[d] Beijing Academy of Quantum Information Sciences, Beijing 100193, China

**ABSTRACT**

Magnesium (Mg) and its alloys have attracted considerable attention owing to their excellent biodegradable properties and biocompatibility. Novel Mg–Sn–Ge ternary Mg alloys were developed as potential biodegradable materials for orthopedic applications because of their alloying elements naturally present in humans. The feasibility of these alloys was investigated in terms of mechanical properties, degradation, cytocompatibility, and hemocompatibility. The hardness and elastic modulus of Mg–2Sn–xGe alloys were improved significantly by increasing the Ge content. Among all the alloys, the Mg–2Sn–3Ge alloy displays outstanding biodegradable properties, as evidenced by the electrochemical tests and hydrogen evolution. The degradation products detected on the corroded alloy surfaces weaken at higher Ge levels. The *in vitro* cytotoxicity assay and hemolysis test showed that the Mg–2Sn–xGe alloys exhibit favorable biocompatibility and hemocompatibility, except for the Mg–2Sn–2Ge alloy.

## 1. Introduction

Magnesium (Mg) and Mg-based alloys are promising bioresorbable implant materials used widely in orthopedic devices. Mg implants provide mechanical support to the human body and naturally dissolve after tissue healing, which avoids the complications and financial loss to the patient resulting from the second surgery to remove the implant [1, 2]. The Mg ions released also play significant functional roles in stimulating bone growth, participating in the enzyme metabolism, and promoting other biological systems [3]. Moreover, Mg and its Mg alloys have superior mechanical properties, and their hardness and elastic modulus are close to those of human bone. This potentially weakens the implant loosening caused by stress shielding. However, the degradation rate of current Mg alloys is too fast, leading to the formation of local alkaline environments and failure of the mechanical integrity before new bone regeneration and tissue healing can occur [4]. Thus, it is essential to enhance the corrosion resistance of Mg alloys for further clinical application.

[1] Corresponding author: Qing Zhao, E-mail address: qzhaoyuping@bit.edu.cn

One of the mainstream methods applied to improve the biological properties of Mg alloys is optimizing the composition design by selecting appropriate alloying elements. These alloying elements affect the mechanical properties, degradation, and biocompatibility of Mg implants because of their different microstructures and standard electrode potential. However, some metallic ions released from Mg implants may have toxic effects during tissue healing. Although metallic ions (aluminum, zinc, and yttrium) improve the corrosion resistance and mechanical properties of Mg alloys, excess amounts are toxic to the human body, potentially resulting in neurotoxicity, Alzheimer's disease, and hepatotoxicity [5]. Therefore, a new series of Mg alloys with essential or trace elements for the human body is highly desired.

Tin (Sn) is one of the most essential trace elements in the human body and regulates many physiological functions. Sn influences the function of hemoglobin and promotes tissue growth and healing [6]. Sn effectively refines the secondary dendrite arm spacing of the α-Mg, improving the mechanical properties [7]. The creep resistance of the Mg–Sn system is preferable to that of AZ alloys because of the precipitated phase ($Mg_2Sn$) [8]. The addition of Sn significantly enhances the corrosion resistance of the Mg–Sn system due to the formation of Sn-contained surface layer [9]. However, there are few reports on the biocompatibility of Mg–Sn alloys. Therefore, Sn was chosen as one of the alloying elements added to Mg implants to evaluate orthopedic biomaterials.

Germanium (Ge) has similar chemical properties to Sn because they are in the same group in the periodic table. Ge is a dietary trace element of the human body and has been reported to regulate many physiological functions [10]. Ge plays important roles in relieving fatigue, preventing anemia, and strengthening the metabolism, and it has anti-tumor and anti-cancer properties [11]. Furthermore, Ge has been reported to be an effective grain refiner for α-Mg dendrites [12]. Bian et al. reported that the strengths and elongations of as-cast Mg–Ge alloys were superior to that of the other as-cast binary Mg alloys. In addition, the Mg–Ge alloys also presented excellent corrosion resistance and biocompatibility [13]. However, the effects of Ge alloying on the biological properties of ternary Mg alloys are not widely reported.

Mg–Sn–Ge ternary alloys have been designed for orthopedic materials owing to the favorable biological properties of Sn and Ge in the Mg matrix. In particular, the influence of the synergy of Ge with Sn on the degradation and biocompatibility of Mg alloys has not been previously studied. Mg–2Sn–xGe (x = 1, 2, or 3 wt.%) alloys were developed, and their feasibility was assessed from the microstructure characterization, mechanical behaviors, biodegradable properties, and biocompatibility aspects for use as orthopedic materials.

## 2. Materials and methods
### 2.1. Material preparation

Mg–Sn–Ge alloys were fabricated by melting high-purity Mg 99.95%, Sn (99.99%), and Ge (99.9999%) in a crucible resistance furnace protected with flux at 700–800°C. The molten metals were poured into a steel mold with a cavity of size ф 30 mm × 100 mm to cool. Table **1** lists the chemical compositions of the Mg–Sn–Ge alloys determined by inductively coupled plasma–optical emission spectrometry (ICP–OES, 725-ES, Agilent, USA). The Mg alloy ingots were processed into disks with dimensions of 10 × 10 × 2 mm$^3$ by wire-cut electrical discharge machining. The disk-shaped samples were mechanically grounded using SiC sandpapers from 600 to 2000 grit, followed by ultrasonic cleaning with acetone, absolute ethanol, and air drying.

The grounded samples were polished into a mirror-like surface with a 1 μm diamond suspension. The samples were then etched with 4% $HNO_3$/alcohol solution for optical microscopy (OM, AXIO Imager M1, Zeiss, Germany) observations. The microstructures of the Mg–2Sn–xGe alloys were examined further by scanning electron microscopy (SEM, S–4800, Hitachi, Japan) coupled with an energy-dispersive X-ray spectroscopy (EDS) system. X-ray diffraction (XRD, Bruker D8 advance, Bruker, Germany) was used to determine the constituent phases precipitated in the alloys operated at 40 kV with 2$\theta$ angles ranging from 20° to 80°.

### 2.2. Nanoindentation test

The nanoindentation equipped with a pyramid Berkovich diamond indenter was used to analyze the mechanical behaviors of the multi-component Mg–Sn–Ge alloys. The samples were polished with a 1 μm diamond suspension because the surface roughness was highly sensitive to the nanoindentation results. The applied load and corresponding displacement were recorded in real-time with a maximum load of 30 mN.

**2.3. Electrochemical test**

The corrosion properties of Mg–Sn–Ge alloys were evaluated from the open circuit potential (OCP), potentiodynamic polarization, and electrochemical impedance spectroscopy (EIS) measurements on an electrochemical workstation (CHI760e, Chenhua Inc., China) in Hank's solution at 37°C. The electrochemical tests were performed using a conventional three-electrode system. The platinum electrode, saturated calomel electrode, and test sample with an exposed area of 1 cm were used as the counter, reference, and working electrodes, respectively. The OCP was continuously recorded for 1 h to achieve a relatively stable potential, followed by the EIS test with the frequency ranging from 100 kHz to 100 mHz at a 5 mV amplitude perturbation. The potentiodynamic polarization test was conducted at a scanning rate of 1 mV/s. The EIS data were fitted using ZView software. The polarization curves were fitted using CView software to derive the corrosion current density ($I_{corr}$) and corrosion potential ($E_{corr}$) from the Tafel regions. At least three repeated tests were performed for each alloy.

**2.4. Hydrogen evolution test**

A hydrogen evolution test was conducted using a device with a calibrated burette and beaker [14] to investigate the long-term corrosion behavior. In line with ASTM G31-72 [15], the samples were immersed into Hank's solution with a ratio (solution volume to sample surface) of 20 mL/cm at 37°C. The volume of hydrogen was recorded at each time point. After immersion for 3 and 20 days, samples were taken from the solution, rinsed with deionized water, and dried in air. The surface morphology and composition of the corroded samples were detected by SEM coupled with EDS. Fourier transform infrared (FTIR) spectroscopy (Nexus 670, Nicolet, USA) was used to determine the functional groups in the corrosion products formed on the corroded sample surfaces, measured in the spectrum range from 4000 to 500 cm$^{-1}$. The chemical compositions of the degradation products were analyzed by X-ray photoelectron spectroscopy (XPS, Axis Ultra, Kratos Analytical Ltd.). High-resolution scanning was carried out to identify the binding energies of Mg 1s, Ca 2p, Na 1s, C 1s, O 1s, and P 2p.

**2.5. Cell viability test**

The Mg–2Sn–xGe Mg alloys were sterilized by ultraviolet radiation for at least 2 hours. The alloys were incubated in Dulbecco's Modified Eagle's Medium (DMEM, Gibco) for 72 h at an extraction ratio of 1.25 cm$^2$/mL in a 5% $CO_2$ incubator at 37°C. The extracts were achieved by filtering the supernatants through a 0.22 μm filter, and stored at 4°C. The ion concentration of the extracts was determined by ICP–OES. MC3T3-E1, the mouse osteoblast cells obtained from American Type Culture Collection, were cultured in the DMEM with 10% fetal bovine serum (FBS), 100 U/mL penicillin, and 100 μg/mg streptomycin in 5% $CO_2$ incubator at 37°C. The cells were seeded in a 96-well plate at a density of 1 × 10 cells per well. After culturing for 24 h to allow cell attachment, the medium was replaced with the alloy extracts. The cells cultured with DMEM were used as the control group. After incubation for 1 and 3 days, a 10 μL MTT solution was added to each well, followed by 4 h of incubation. Subsequently, 150 μL of dimethyl sulfoxide was added to dissolve the formed formazan crystals from MTT. The spectrophotometric absorbance of each well at 490 nm was measured using a microplate reader (Cytation 3, BioTek, USA). The cell viability of the MC3T3-E1 cells of the Mg–2Sn–xGe alloys was calculated using the following equation:

$$\text{Viability} = \frac{\text{OD}_{\text{sample}}}{\text{OD}_{\text{control}}} \times 100\% \quad (1)$$

**2.6. Hemolysis test**

Healthy mouse blood was anticoagulated using 3.8 wt. % sodium citrate at a 9:1 ratio and diluted with phosphate-buffered saline (PBS) at a volume ratio of 4:5. The alloy extracts were prepared in a centrifuge tube containing 14 mL PBS at 37°C for 30 min. Subsequently, 0.2 mL of diluted blood was added into each centrifuge tube and cultured at 37°C for another 60 min. PBS was used as the negative group, and deionized water as the positive group. After incubation, all the tubes were centrifuged at 3000 rpm for 5 min. The supernatant was pipetted carefully into a 96-well plate and measured using a microplate reader at 545 nm. The hemolysis rate was calculated based on the results of three replicates using the following equation:

$$\text{Hemolysis rate} = \frac{\text{OD}_{\text{sample}} - \text{OD}_{\text{negative}}}{\text{OD}_{\text{positive}} - \text{OD}_{\text{negative}}} \times 100\% \quad (2)$$

## 3. Results

### 3.1. Microstructural analysis

Fig. **1** presents the XRD patterns of the as-cast Mg–2Sn–xGe alloys. The Mg–2Sn–xGe alloys mainly consist of α–Mg matrix and Mg2Ge phase with a face-centered cubic crystal system. The Mg–2Sn–xGe alloys show a relatively weak XRD peak for the $Mg_2Sn$ intermetallic phase.

Fig. **2**(a) shows the optical microstructures of the Mg–2Sn–xGe alloys. The Mg–2Sn–xGe alloys exhibit a coarse dendritic α–Mg phase and a continuous network eutectic $Mg_2Ge$ phase. The grain size of the α–Mg dendrites became refined as the Ge content was increased. Furthermore, almost no $Mg_2Sn$ precipitated on Mg–2Sn–xGe alloys, which is consistent with the XRD patterns with a weak peak for the $Mg_2Sn$ phase. Fig. **2**(b-c) presents SEM images at different magnifications to provide further insight into the phase distribution. Table **2** lists the corresponding EDS results of points labeled in Fig. **2**(c). The chemical composition of the gray matrix (Position A) in the Mg–2Sn–1Ge alloy is only Mg, corresponding to the α-Mg matrix. The white rod-like precipitate (Position B) is rich in Mg and Ge, containing 69.09 at.% (Mg) and 30.91 at.% (Ge), respectively, indicating the formation of $Mg_2Ge$ intermetallic compounds. More white rod-like precipitates are observed as the Ge content is increased [12]. Position C contains a large amount of Mg and Sn and a small amount of Ge at concentrations of 80.03 at.% (Mg), 18.62 at.% (Sn), and 1.35 at.% (Ge), respectively, which is related to the $Mg_2Sn$ and $Mg_2Ge$ phases. Similar EDS results are found in the Mg–2Sn–2Ge and Mg–2Sn–3Ge alloys.

### 3.2. Mechanical properties

Fig. **3**(a) presents the load-displacement curves of the Mg–2Sn–xGe Mg alloys. In the loading process, the surfaces of the samples initially present elastic deformation and then exhibit plastic deformation as the load intensifies. The maximum indentation depth of Mg–2Sn–3Ge alloy is 1159.9 ± 5.8 nm, which is significantly less than that of the Mg–2Sn–2Ge (1314.3 ± 14.5 nm) and Mg–2Sn–1Ge (1368.7 ± 8.4 nm) alloys. The elastic recovery is 16.3% for Mg–2Sn–1Ge, 17.0% for Mg–2Sn–2Ge, and 18.1% for Mg–2Sn–3Ge, suggesting that the plastic deformation resistance is enhanced as the Ge content is increased [16]. Fig. **3**(b) shows the average hardness and elastic modulus values of samples. Both the hardness and modulus are enhanced significantly as the Ge content is increased. Hence, adding Ge can improve the surface mechanical properties of samples.

### 3.3. Electrochemical evaluations

The OCP curves of Mg–2Sn–xGe alloys in Hank's solution are monitored (Fig. **4**(a)). The OCP values of the samples fluctuate considerably before soaking for 400 s, resulting from the formation and subsequent dissolution of the degradation product (Mg(OH)$_2$) in Hank's solution containing Cl$^-$ ions [17, 18]. Subsequently, the steady OCP curves are observed, indicating the stable formation and dissolution of Mg(OH)$_2$. The Mg–2Sn–3Ge alloy shows a slight increase in the steady OCP value compared to other alloys. Fig. **4**(b) shows the potentiodynamic polarization curves of the Mg–2Sn–xGe alloys. The corrosion potential ($E_{corr}$) moves in the positive direction with increasing Ge content. The increased addition of Ge inhibits the anodic active dissolution [19]. The Mg–2Sn–3Ge alloy displays an inferior cathodic current density compared to the other alloys, indicating the suppressed hydrogen evolution reaction. Table **3** lists the electrochemical parameters derived from the polarization curves. The Mg–2Sn–3Ge alloy exhibits the lowest $I_{corr}$ and highest $E_{corr}$, followed by the Mg–2Sn–2Ge alloy, indicating that the corrosion resistance increased as the Ge content is increased.

Fig. **5** presents the EIS spectra of Mg–2Sn–xGe alloys. The Nyquist plots of the Mg–2Sn–xGe alloys (Fig. **5**(a)) are composed mainly of an inductive loop at low frequency and two capacitance loops at medium and high-frequency domains. The occurrence of an inductive loop is relevant to ion adsorption [20]. The two capacitive loops are assigned to the degradation product layer and charge transfer, respectively. The increase in the semi-circle size became increasingly significant as the Ge concentration is increased [21]. As shown in Fig. **5**(b), the impedance modulus increases dramatically at high and medium frequencies and then decreases at low frequencies. In particular, Mg–2Sn–3Ge possesses the highest impedance at a low frequency, indicating superior corrosion resistance among the alloys. Fig. **5**(c) shows the Bode phase angle plots. Two wave crests corresponding to the two capacitive loops and one wave trough representing the inductance loop are observed. A relatively broad peak above 45° is observed for the Mg–2Sn–3Ge alloy, confirming the improved corrosion resistance with Ge content. Therefore, an increase in Ge content improves the corrosion resistance of the Mg–2Sn–xGe alloy system, which leads to the enlarged Nyquist loop size, increased Bode impedance, and accrescent phase angle. These results are consistent with the potentiodynamic polarization.

The degradation characteristics of the Mg–2Sn–xGe alloys were examined using the equivalent circuit in Fig. 6 to fit the EIS spectra. The solution resistance is expressed by $R_s$, which depends on the conductivity of the tested solution. The constant phase element ($CPE_f$) is in parallel with the resistance $R_f$, which is relevant to the degradation products. $CPE_{dl}$ denotes the constant phase element for the double-layer capacitance at the electrolyte/substrate interface [22]. $R_t$ is the corresponding charge transfer resistance. L and $R_L$ correspond to the inductance and the inductance resistance, indicating the existence of pitting corrosion at a low frequency. Table **4** lists the parameters of the equivalent circuit fitted using ZView software.

Typically, the CPE, instead of an ideal capacitor, is to investigate the inhomogeneous surface due to cracks, impurities, and secondary phases [20]. The components of CPE are Q and n ($0 \leq n \leq 1$). The CPE stands for pure resistance or capacitance when n = 0 or 1, respectively. While, the value of Q cannot denote the capacitance exactly when n < 1, as reported by Orazem et al. [23, 24]. The relationship between the CPE elements and effective capacitance $C_{eff}$ is expressed using the following equation [25]:

$$C_{eff} = Q^{1/n_f}\left(\frac{1}{R_s} + \frac{1}{R_f}\right)^{(n_f-1)/n_f} \quad (3)$$

The variation of $C_{eff}$ is inversely proportional to the thickness of the product layer, d [26]:

$$d = \frac{\varepsilon\varepsilon_0}{C_{eff}} \quad (4)$$

where $\varepsilon$ and $\varepsilon_0$ represent the dielectric constant and vacuum permittivity, respectively. According to Eq. 3, the $C_{eff}$ of the Mg–2Sn–xGe alloys are 93.12 µF/cm$^2$ (Mg–2Sn–1Ge), 35.87 µF/cm$^2$ (Mg–2Sn–2Ge), and 26.48 µF/cm$^2$ (Mg–2Sn–3Ge), respectively. This confirms that the degradation product layer increases with increasing of Ge content during the initial time.

The polarization resistance ($R_p$) [19] is expressed by Eq. 5:

$$R_p = R_f + \left(\frac{1}{R_t} + \frac{1}{R_L}\right)^{-1} \quad (5)$$

The enlarged resistance ($R_f$, $R_t$, and $R_p$) contribute to better protection of the alloys from aggressive ions. Compared to $R_f$ in the Mg–2Sn–1Ge alloy, $R_t$ displays a more than 10-fold increase, indicating that the charge transfer process plays a leading role in hindering the penetration of ions. A similar variation is observed in the remaining alloys. Moreover, Mg–2Sn–3Ge alloy achieves the maximum $R_f$, $R_t$, and $R_p$ values, suggesting the minimum degradation rate.

### 3.4. Hydrogen evolution

Fig. **7** displays the variation of hydrogen evolution volume during immersion time. At the initial soaking stage, a smaller difference in hydrogen evolution rates is observed among tested alloys. While the evolved hydrogen of Mg–2Sn–1Ge alloy increases continuously with the prolongation of immersion, exhibiting a high hydrogen evolution rate. Before immersion for 150 h, both Mg–2Sn–2Ge and Mg–2Sn–3Ge possess relatively lower hydrogen evolution rate, suggesting that their thicker product layer hinders the corroded ions from penetrating. Subsequently, the increase in hydrogen evolution rate in Mg–2Sn–2Ge is greater than that in the Mg–2Sn–3Ge alloy. This is attributed to the rupture of the degradation product layer on the surface of Mg-2Sn–2Ge alloy. By contrast, the hydrogen evolution rate of the Mg–2Sn–3Ge alloy remains at the lowest level throughout the immersion process, suggesting that the high content can delay the crack of the product layer and enhance the anti-corrosion behavior.

Fig. **8** shows the surface morphologies and the element contents detected by EDS of Mg–2Sn–xGe alloys soaked in Hank's solution for 3 days and 20 days. After immersion for 3 days, a degradation product layer with some particle/globosity deposition is observed on the surfaces of Mg–2Sn–xGe alloys. The visible cracks spread over the whole surface of the Mg–2Sn–1Ge alloy, which provides a channel for $Cl^-$ and other ions to penetrate. However, fewer cracks emerged on the Mg–2Sn–(2–3)Ge alloys, suggesting a restrained degradation rate with increasing Ge content. After immersion for 20 days, many degradation products aggregate on Mg–2Sn–xGe alloy surfaces. These degradation products are composed mainly of Mg, C, and O, as shown in the EDS results, suggesting the formation of hydroxyls and carbonates. In particular, the degradation product layer on the Mg–2Sn–1Ge alloy is thick and dense. With increasing Ge content, the degradation degree is suppressed significantly, resulting from the continuous and uniform $Mg_2Ge$ intermetallic compound.

FTIR and XPS spectroscopy are used to determine the chemical composition of the degradation products formed on the Mg–2Sn–xGe alloys surfaces after immersion in Hank's solution for 20 days (Fig. **9**). Since the XPS spectra of the Mg–2Sn–xGe alloys are similar to each other, high resolution narrow scanning of Mg 1s, Ca 2p, Na 1s, O 1s, and P 2p are presented for the Mg–2Sn–1Ge alloy. The FTIR spectra reveal the existence of varied functional groups, such as $OH^-$ [27], $CO_3^{2-}$ [28], and $PO_4^{3-}/HPO_4^{2-}$ [29]. Furthermore, XPS is used to confirm the surface composition of the corresponding compounds. The C 1s spectra can be split into three peaks: 284.8 eV (C–C/C–H), 286.1 eV (C–O), and 289.7 eV (C=O) [30]. The presence of $CaCO_3$ can be confirmed by the binding energy of Ca 2p spectra at 351.5 eV along with C 1s peak at 289.7 eV and O 1s peak at 532.5 eV [13]. In addition to the O 1s peak at 532.5 eV, the peaks of O 1s at 531.3 eV and Mg 1s peak at 1303.8 eV are assigned to the $Mg(OH)_2$. The binding energy of P 2p stands at 133.3 eV, which is associated with the presence of phosphates [31]. The FTIR coupled with XPS spectra indicate that a degradation product layer mainly consisted of hydroxide, carbonate, and phosphate is formed after soaking for 20 days.

### 3.5. Biocompatibility evaluation

Fig. **10**(a) presents the viability of MC3T3-E1 cells cultured in Mg–2Sn–xGe alloys extracts for 1 and 3 days. All the Mg–2Sn–xGe alloys show increased cell viability exceeding 80% during incubation for 1 day, suggesting no significant toxic effects followed by the ISO 10993-5 [32]. However, the cell viability in the three alloys changes considerably after culturing for 3 days. Except for the Mg–2Sn–3Ge alloy, the viability of MC3T3-E1 cells incubated in the extracts of the remaining alloys decreases with time, implying a hysteretic response of MC3T3-E1 cells. In particular, the Mg–2Sn–2Ge alloy displays a cell viability of slightly less than 80%, indicating a degree of toxicity. On the other hand, the viability in the Mg–2Sn–3Ge alloy extract is higher than that in other extracts throughout the incubation period, indicating good biocompatibility. The cell activity is related to the release of metal ions in the extract. Fig. **10**(b) shows the metal ion concentrations of the corresponding extracts in Mg–2Sn–xGe alloys. The Mg–2Sn–2Ge alloy releases slightly higher concentrations of ions than that of Mg–2Sn–1Ge, and the Mg–2Sn–3Ge alloy has the lowest ion content. The rapid degradation of samples increases the concentration of metal ions in the extracts, which inhibits cell proliferation. Therefore, the Mg–2Sn–3Ge alloy achieves excellent viability with MC3T3-E1 cells because of the low ion concentrations in extracts.

Fig. **11** shows the hemolysis rates of Mg–2Sn–xGe alloys. The Mg–2Sn–2Ge alloy exhibits a high hemolysis rate of 13.68% ± 1.06%. However, the hemolysis rates of the remaining alloys are all lower than 5%, suggesting superior hemocompatibility

[33]. The performance of hemocompatibility for biomaterials depends largely on the degradation behavior, ion release, and pH [34]. The lower hemolysis rates of Mg–2Sn–(1,3)Ge alloys are possibly due to the reduced degradation rate and released ion concentration in the initial stage.

## 4. Discussion

### 4.1. Microstructure and mechanical behavior

In the Mg–2Sn–xGe alloy system, there are relatively weak peaks of the $Mg_2Sn$ phase from the XRD patterns. This may be due to the low concentration of Sn addition. According to the Mg–Sn phase diagram [35], the maximum solid solubility of Sn in Mg is 3.35 at.% (14.48 wt.%) at 561.2°C. The small 2 wt.% Sn dissolves readily in the solid solution, suggesting no obvious $Mg_2Sn$ phase in the Mg–2Sn–xGe alloys. By contrast, Ge has no solid solubility in Mg conformed to the Mg–Ge phase diagram [36], which means that the $Mg_2Ge$ phase is observed as precipitation surrounding the dendritic α-Mg matrix. Therefore, the effects of solid solution strengthening on the mechanical properties of the Mg–2Sn–xGe alloys can be ignored. The intermetallic compound $Mg_2Ge$ presents a network morphology, as shown in Fig. **2**. This network structure becomes more dense and continuous with the increased addition of Ge, resulting in improved hardness. Moreover, intermetallic phase strengthening and grain refinement significantly restrict dislocation mobility, contributing to the improvement of plastic deformation resistance [37]. Therefore, the mechanical properties of the Mg–2Sn–xGe alloys improve with the increased Ge addition.

### 4.2. Biodegradable properties

The standard potentials of the alloying elements are −2.77 V (Mg), −0.14 V (Sn), and 0.24 V (Ge), respectively [38, 39]. Ge possesses a nobler electrode potential than Sn. Therefore, Ge enhances the electrode potential between Sn and the α-Mg matrix in the tested alloys. The corrosion potential shifts to positive values as the Ge content is increased, as shown in Fig. **4**(b) and Table **3**. Visible $Mg_2Ge$ and fewer $Mg_2Sn$ second phases are observed on the Mg–2Sn–xGe alloys, as shown in Figs. **1-2**. Hence, the difference in the degradation behavior of Mg–2Sn–xGe alloys is mainly due to the change in the microstructure and volume fraction of the $Mg_2Ge$ phase. The second $Mg_2Ge$ phase deposited in Mg–2Sn–xGe alloys normally plays a dual role during degradation [39, 40]. On the one hand, a micro-battery is formed between the second phase and α-Mg matrix because of the difference in corrosion potential. The Mg$_2$Ge phase is acted as the cathode electrode due to its higher potential relative to the matrix phase, accelerating the degradation process because of galvanic corrosion. On the other hand, the second phase provides a barrier against corrosion and reduces the degradation rate. The Mg–2Sn–1Ge alloy, with a lower volume fraction of the Mg$_2$Ge phase, cannot protect the α-Mg matrix facilitating local evolution due to galvanic corrosion. This is highlighted by the surface morphology with distinct cracks and thick degradation products (Fig. **8**). However, with increasing Ge content, a continuous reticular Mg$_2$Ge phase is formed on the sample surfaces, as shown in Fig. **2**, which effectively prevents the degradation of the matrix from aggressive ions. Therefore, the continuous network of the Mg$_2$Ge phase acts as a barrier to the matrix leading to retard degradation of Mg–2Sn–3Ge alloy.

A degradation process involves with thermodynamic and kinetic factors, such as passive film and corrosion current density [41]. Mg dissolves quickly when the Mg–2Sn–xGe alloys are soaked in the physiological Hank's solution. The oxidation–reduction process can be expressed as follows:

$$\text{Anodic reaction: Mg} \rightarrow \text{Mg}^{2+} + 2e^- \tag{6}$$

$$\text{Cathodic reaction: } 2H_2O + 2e^- \rightarrow 2OH^- + H_2 \tag{7}$$

$$\text{Overall reaction: Mg} + 2H_2O \rightarrow Mg(OH)_2 + H_2 \tag{8}$$

The low value of $R_t$ in Mg–2Sn–1Ge alloy indicates accelerated degradation, resulting in visible dense cracks on the surface. However, the degradation process is alleviated as the Ge content is increased because of the continuous and reticular Mg$_2$Ge phase deposited on the surface. This suggests that the barrier effect is dominant relative to the micro-battery effect in samples with high Ge content. Mg dissolves at the anode to produce Mg synchronized with water reduction reaction at cathodic producing OH during the degradation process [17]. Thus, insoluble Mg(OH)$_2$ is formed and covers the surfaces of the samples serving as a protective layer for the substrates. The corrosion product Mg(OH)$_2$ combined with the reticular Mg$_2$Ge deposits provide efficient protection from ion attacks for the Mg–2Sn–3Ge alloy, which is consistent with the results of a larger value of $R_f$ in EIS. As the immersion time is increased, the deposited layer of degradation products dissolves due to the attack of aggressive Cl$^-$ ions:

$$\mathrm{Mg(OH)_2 + 2Cl^- \rightarrow MgCl_2 + 2OH^-} \tag{9}$$

As the degradation intensified, more degradation products are formed on the sample surfaces, mainly including oxide, hydroxide, carbonate, and phosphate. These products can be degraded and absorbed metabolically without causing side effects to the human body [42].

**4.3. Biocompatibility of Mg–2Sn–xGe alloys**

A new biodegradable material exhibits appropriate degradation behavior, mechanical properties, and biocompatibility. Both Sn and Ge are essential trace elements for the human body. Sn promotes tissue growth by stimulating the reactions of protein and nucleic acid. Sn can inhibit iron absorption and promote hemoglobinolysis, affecting the function of hemoglobin [6]. Sn is also an inducer of renal heme oxygenase and a participant in energy metabolism [43]. Ge is conducive to many physiological functions, such as improving immunity, stimulating circulation, protecting red blood cells, and reducing high blood pressure [10]. Moreover, Ge can be absorbed or excreted through the metabolic system. Generally, the average Sn and Ge intake in the human diet are 2–3.5 and 0.4–3.4 mg per day, respectively [10, 44, 45].

The concentration of metal ions in alloy extracts is responsible for the cytotoxicity. The higher contents of ions in the extracts restrain the cell viability, but lower ion concentrations enhance cell proliferation. A previous study reported that 15 mM (360 μg/mL) Mg is not toxic to MC3T3-E1 cells [46]. The highest metal ion concentrations (Mg: 341.58 ± 17.70 μg/mL, Sn: 0.13±0.07 μg/mL, Ge: 4.52±0.15 μg/mL) are observed in the Mg–2Sn–2Ge extracts, indicating reduced cell viability. The Mg–2Sn–xGe alloys are immersed in the DMEM to prepare the extracts.

Mg–2Sn–3Ge alloy has minimal degradation (i.e., the lowest concentration of released ions) because of the continuous and dense network $Mg_2Ge$ deposits, resulting in excellent cell proliferation. However, as to the Mg–2Sn–2Ge alloy, the degradation behavior is not stable during the 3-day extraction period. The limited $Mg_2Ge$ deposits cannot act effectively as a barrier during the short period of extraction, leading to the release of maximum metallic ions (as shown in Fig. 10(b)). From this point, the Mg–2Sn–3Ge alloy achieves good biocompatibility because of its superior stability and degradability.

The rupture of red blood cells (RBC) by Mg implants is caused mainly by mechanical damage from the alloy surface and different osmotic pressures between the inside and outside of the RBC from the release of metal ions [39, 47]. The shape, thickness, and roughness of all the samples in the hemolysis assay are identical. Thus, the mechanical damage to the RBC resulting from the alloy surface can be ignored. The hemocompatibility of the Mg–2Sn–xGe alloys is mainly due to the concentration of released metal ions in the degradation process. The Mg–2Sn–2Ge alloy exhibits unsatisfied hemocompatibility, indicating a rapid degradation rate. This is demonstrated by the increased concentration of ions in Fig. 10(b).

## 5. Conclusions

New biodegradable ternary Mg alloys are designed, and their potential for orthopedic applications is evaluated systematically. The Mg–2Sn–xGe alloys are mainly composed of α-Mg matrix and $Mg_2Ge$ phase. No obvious $Mg_2Sn$ phase is observed because of the low Sn content and highly solid solubility of Sn in Mg. The network structure of the $Mg_2Ge$ phase grows continuous and dense with increasing Ge content. The degradation performance of the Mg–2Sn–xGe alloys is investigated using electrochemical tests, hydrogen evolution, and degradation product analysis. The lowest degradation rate is observed with Ge contents up to 3.0 wt.%. This is attributed to the dominant barrier effects of intermetallic compounds over the galvanic corrosion effects with increasing Ge content. Regarding to biocompatibility, the Mg–2Sn–3Ge alloy exhibits excellent cell viability and cell adhesion among the tested alloys. The Mg–2Sn–xGe alloys also exhibits adequate hemocompatibility, except for Mg–2Sn–2Ge. Considering the degradation behavior, cytotoxicity, and hemocompatibility, the Mg–2Sn–3Ge alloy is recommended for further *in vivo* studies for orthopedic applications.

**Declaration of Competing Interest**

The authors declare that they have no known competing financial interests or personal relationships that could have appeared to influence the work reported in this paper.

**CRedit authorship contribution statement**

**Xian Wei**: Methodology, Investigation, Writing the original manuscript. **Sujie Ma**: Investigation, Formal analysis. **Jiajia Meng**: Methodology, Validation. **Hong Qing**: Data curation, Validation. **Qing Zhao**: Conceptualization, Funding acquisition.

**Acknowledgments**

This work was supported by the National Science Foundation (NSF) of China (grant No. 11675014), and the Ministry of Science and Technology of China

(2013YQ030595-3). Additional support was provided by Fundamental Research Program of Shanxi Province in China (20210302124264).

Table 1 Chemical compositions of the Mg-Sn-Ge Mg alloys (in wt. %).

| Alloys | Chemical compositions (in wt. %) | | | | | | | | |
|---|---|---|---|---|---|---|---|---|---|
| | Sn | Ge | Na | Ca | Al | Si | P | Mn | Mg |
| Mg-2Sn-1Ge | 2.089 | 1.028 | 0.029 | 0.027 | 0.022 | 0.014 | 0.010 | 0.006 | Bal. |
| Mg-2Sn-2Ge | 1.873 | 2.091 | 0.028 | 0.035 | 0.026 | 0.037 | 0.010 | 0.006 | Bal. |
| Mg-2Sn-3Ge | 2.173 | 3.134 | 0.014 | 0.031 | 0.021 | 0.019 | 0.003 | 0.006 | Bal. |

Table 2 The EDS results of points labeled in Fig. 2(c)

| Alloys | Position | Mg (at. %) | Ge (at.%) | Sn (at.%) |
|---|---|---|---|---|
| Mg-2Sn-1Ge | A | 100.00 | 0 | 0 |
| | B | 69.09 | 30.91 | 0 |
| | C | 80.03 | 1.35 | 18.62 |
| Mg-2Sn-2Ge | D | 65.92 | 32.26 | 1.82 |
| | E | 80.84 | 3.41 | 15.75 |
| Mg-2Sn-3Ge | F | 66.65 | 31.97 | 1.68 |
| | G | 80.48 | 1.51 | 18.01 |

Table 3 The fitted parameters from the potentiodynamic polarization curves.

| Alloys | $E_{corr}$ (V vs. SCE) | $I_{corr}$ (μA/cm$^2$) | $β_c$ (mV) |
|---|---|---|---|
| Mg-2Sn-1Ge | -1.52 | 310.33 | -326.09 |
| Mg-2Sn-2Ge | -1.51 | 189.69 | -318.01 |
| Mg-2Sn-3Ge | -1.50 | 75.63 | -331.89 |

Table 4 The parameters of the equivalent circuit fitted based on the EIS spectra.

| Alloys | $R_s$ (Ω cm$^2$) | $Q_f$ (Ω$^{-1}$ cm$^{-2}$ S$^n$) | $n_f$ | $R_f$ (Ω cm$^2$) | $Q_t$ (Ω$^{-1}$ cm$^{-2}$ S$^n$) | $n_t$ | $R_t$ (Ω cm$^2$) | $L$ (H cm$^{-2}$) | $R_L$ (Ω cm$^2$) | $R_P$ (Ω cm$^2$) | $χ^2$ (× 10$^{-3}$) |
|---|---|---|---|---|---|---|---|---|---|---|---|
| Mg-2Sn-1Ge | 17.47 | 2.87×10$^{-5}$ | 0.70 | 35.43 | 4.67×10$^{-5}$ | 0.84 | 417.10 | 1071 | 456.60 | 253.41 | 2.64 |
| Mg-2Sn-2Ge | 11.36 | 2.61×10$^{-5}$ | 0.66 | 51.70 | 8.49×10$^{-5}$ | 0.74 | 583.30 | 1360 | 895.90 | 404.98 | 3.08 |
| Mg-2Sn-3Ge | 12.26 | 1.58×10$^{-5}$ | 0.68 | 81.73 | 5.02×10$^{-5}$ | 0.80 | 1020.00 | 5133 | 1048.00 | 598.64 | 4.34 |

**Figure captions**

Fig. 1 XRD patterns of the Mg-2Sn-xGe alloys.

Fig. 2 Microstructures of the Mg-2Sn-xGe alloys: (a) optical micrographs, and (b-c) SEM images.

Fig. 3 (a) Indentation load-displacement curves, and (b) hardness and elastic modulus values of the Mg-2Sn-xGe alloys.

Fig. 4 (a) OCP curves, and (b) potentiodynamic polarization curves of the Mg-2Sn-xGe alloys in Hank's solution.

Fig. 5 (a) Nyquist plots, (b) Bode plots of impedance vs. Frequency, and (c) Bode plots of phase angle vs. Frequency of the Mg-2Sn-xGe alloys. The EIS spectra plotted with scatter and the corresponding fitting curves plotted with solid line.

Fig. 6 Equivalent circuits for the Mg-2Sn-xGe alloys based on EIS data.

Fig. 7 The released volume of hydrogen during the immersion time for Mg-2Sn-xGe alloys.

Fig. 8 Surface morphology and the EDS composition results detected on the corroded surface of the degradation products at immersion time of 3 and 20 days.

Fig. 9 (a) FTIR spectra and (b) XPS analysis of the corroded surface for the Mg-2Sn-xGe alloys after soaking for 20 days; (c-h) high-resolution XPS spectra of Mg 1s, Ca 2p, Na 1s, C 1s, O 1s, and P 2p for the Mg-2Sn-1Ge alloy.

Fig. 10 (a) The viability of MC3Te3-E1 cells and (b) ions concentrations of the extracts from the Mg-2Sn-xGe alloys.

Fig. 11 Hemolysis rates of the Mg-2Sn-xGe alloys.

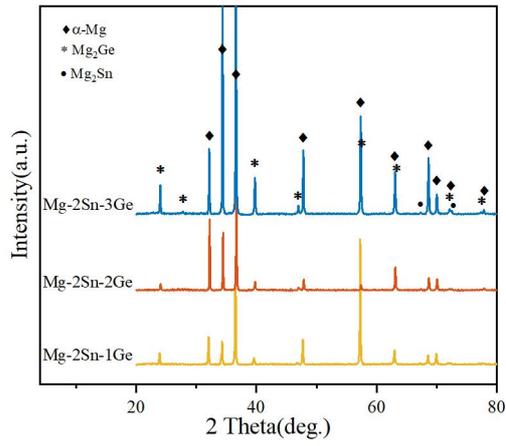

Fig. 1 XRD patterns of the Mg-2Sn-xGe alloys.

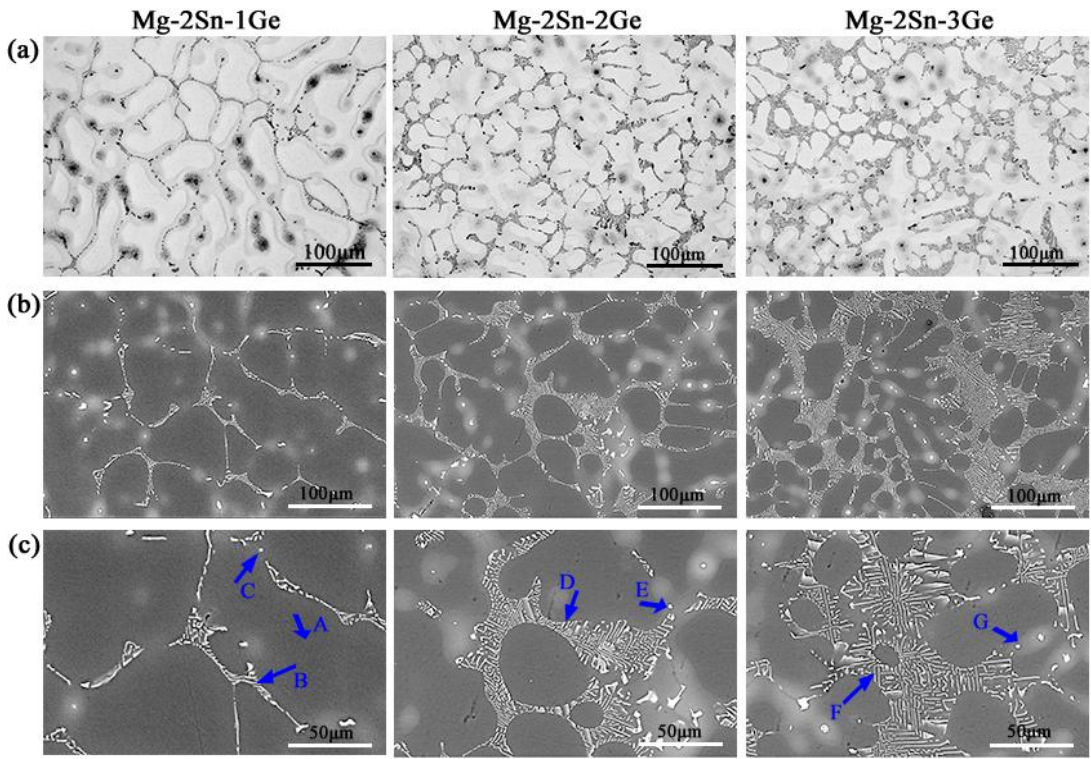

Fig. 2 Microstructures of the Mg-2Sn-xGe alloys: (a) optical micrographs, and (b-c) SEM images.

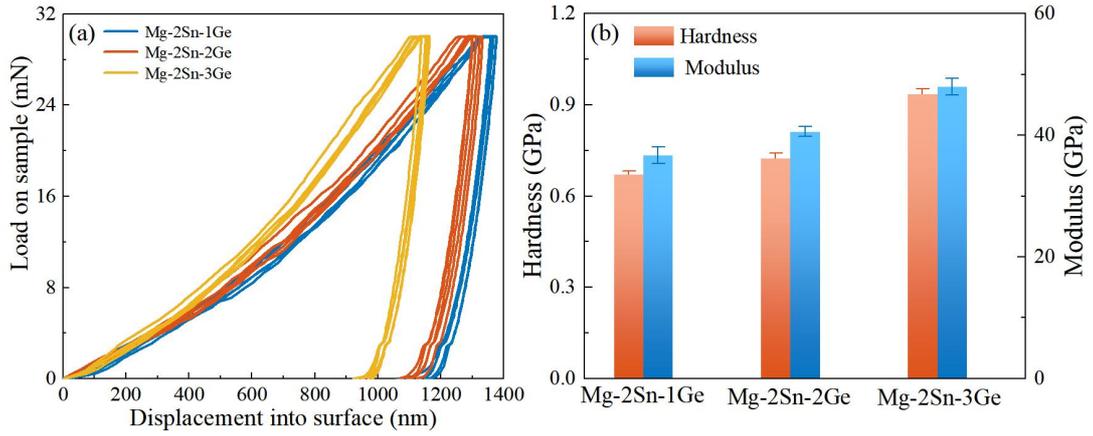

Fig. 3 (a) Indentation load-displacement curves, and (b) hardness and elastic modulus values of the Mg-2Sn-xGe alloys.

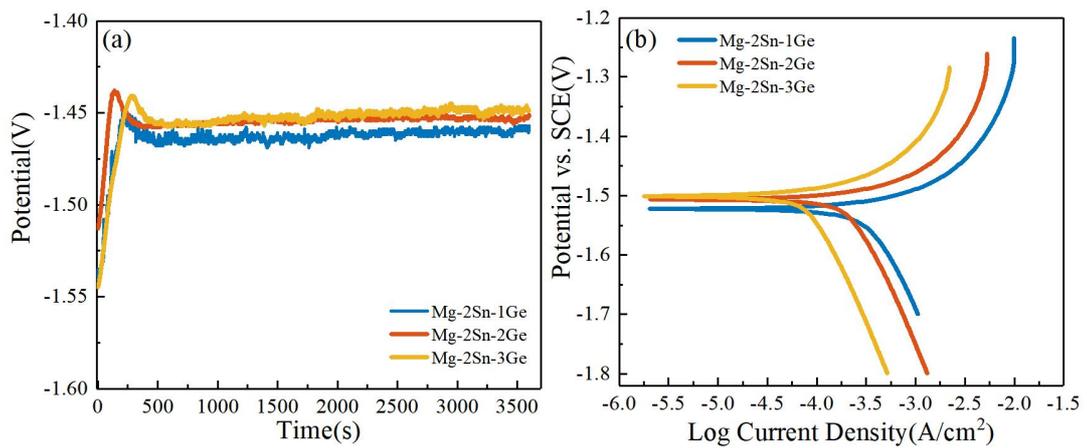

Fig. 4 (a) OCP curves, and (b) potentiodynamic polarization curves of the Mg-2Sn-xGe alloys in Hank's solution.

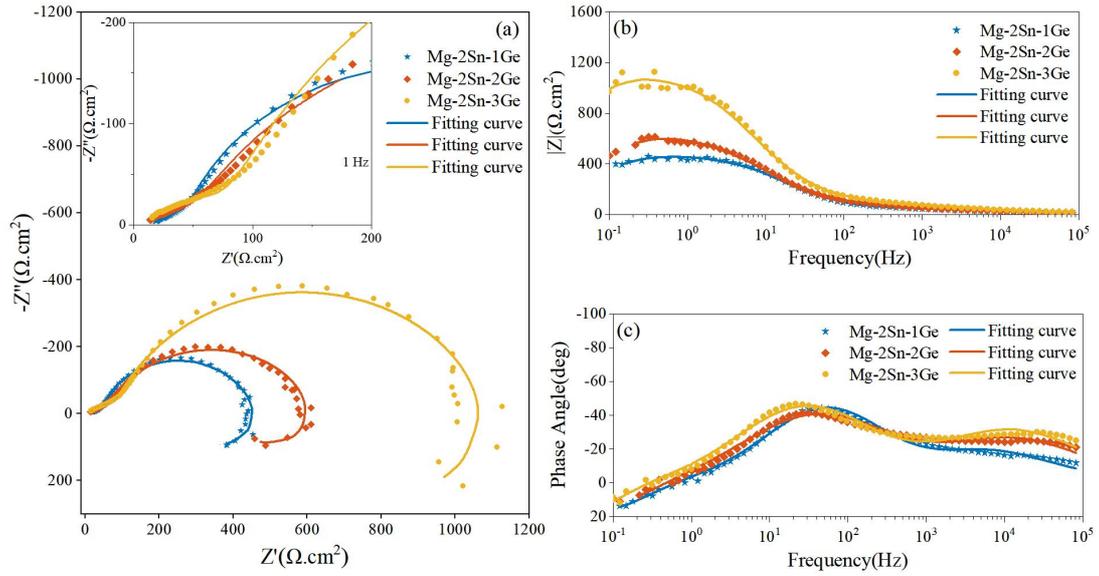

Fig. 5 (a) Nyquist plots, (b) Bode plots of impedance vs. Frequency, and (c) Bode plots of phase angle vs. Frequency of the Mg-2Sn-xGe alloys. The EIS spectra plotted with scatter and the corresponding fitting curves plotted with solid line.

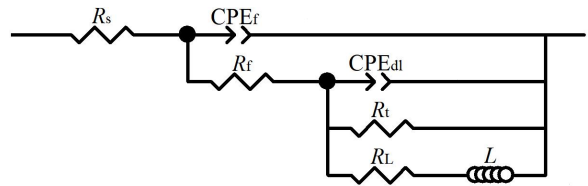

Fig. 6 Equivalent circuits for the Mg-2Sn-xGe alloys based on EIS data.

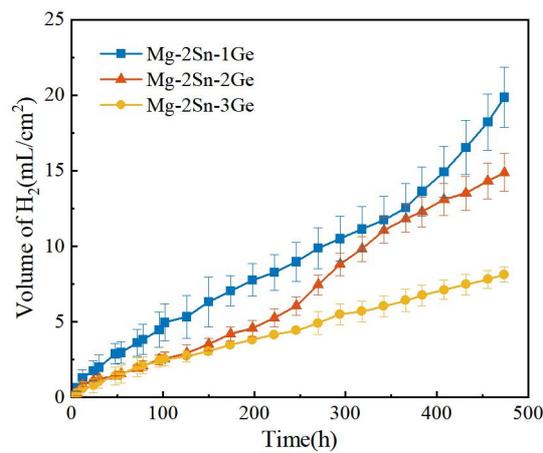

Fig. 7 The released volume of hydrogen during the immersion time for Mg-2Sn-xGe alloys.

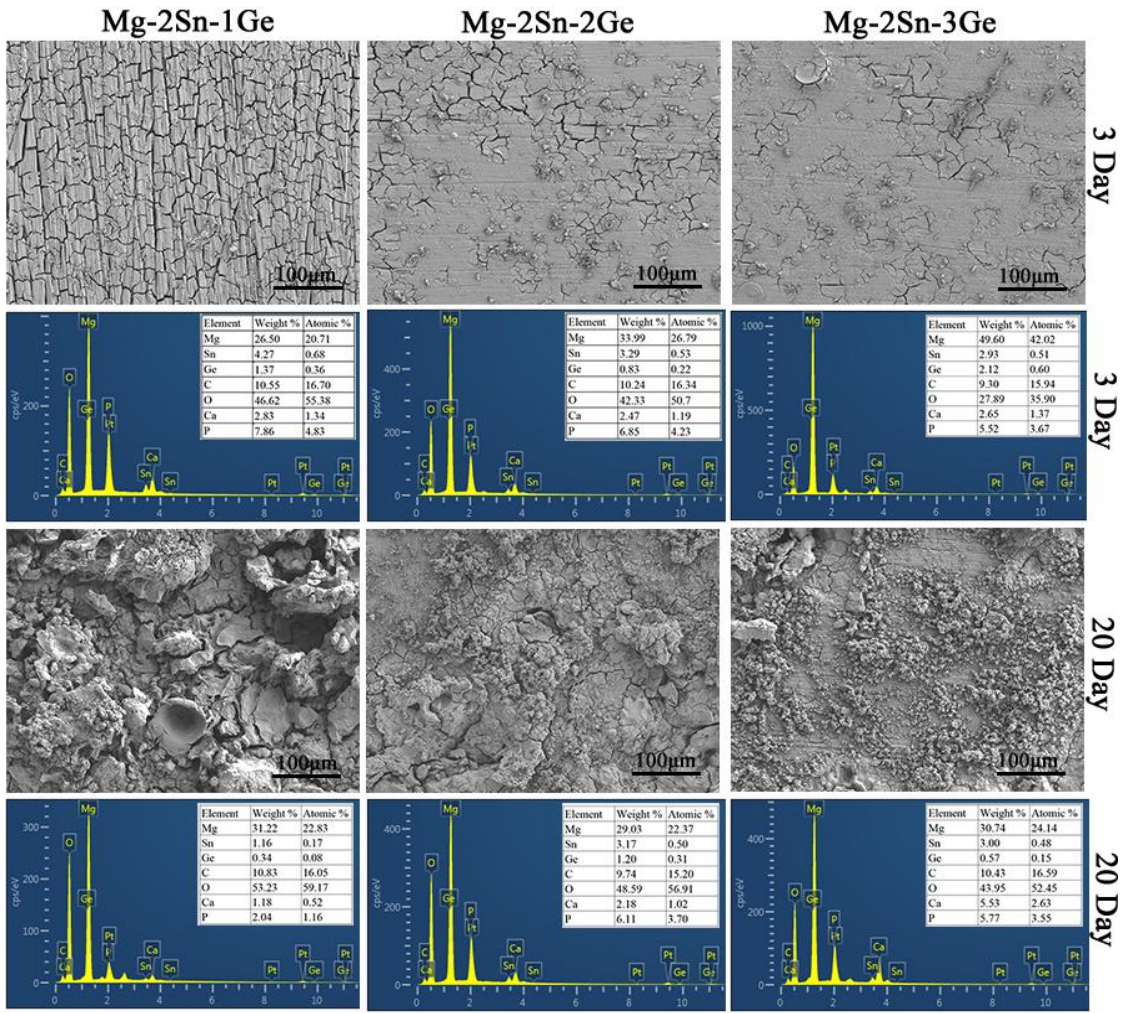

Fig. 8 Surface morphology and the EDS composition results detected on the corroded surface of the degradation products at immersion time of 3 and 20 days.

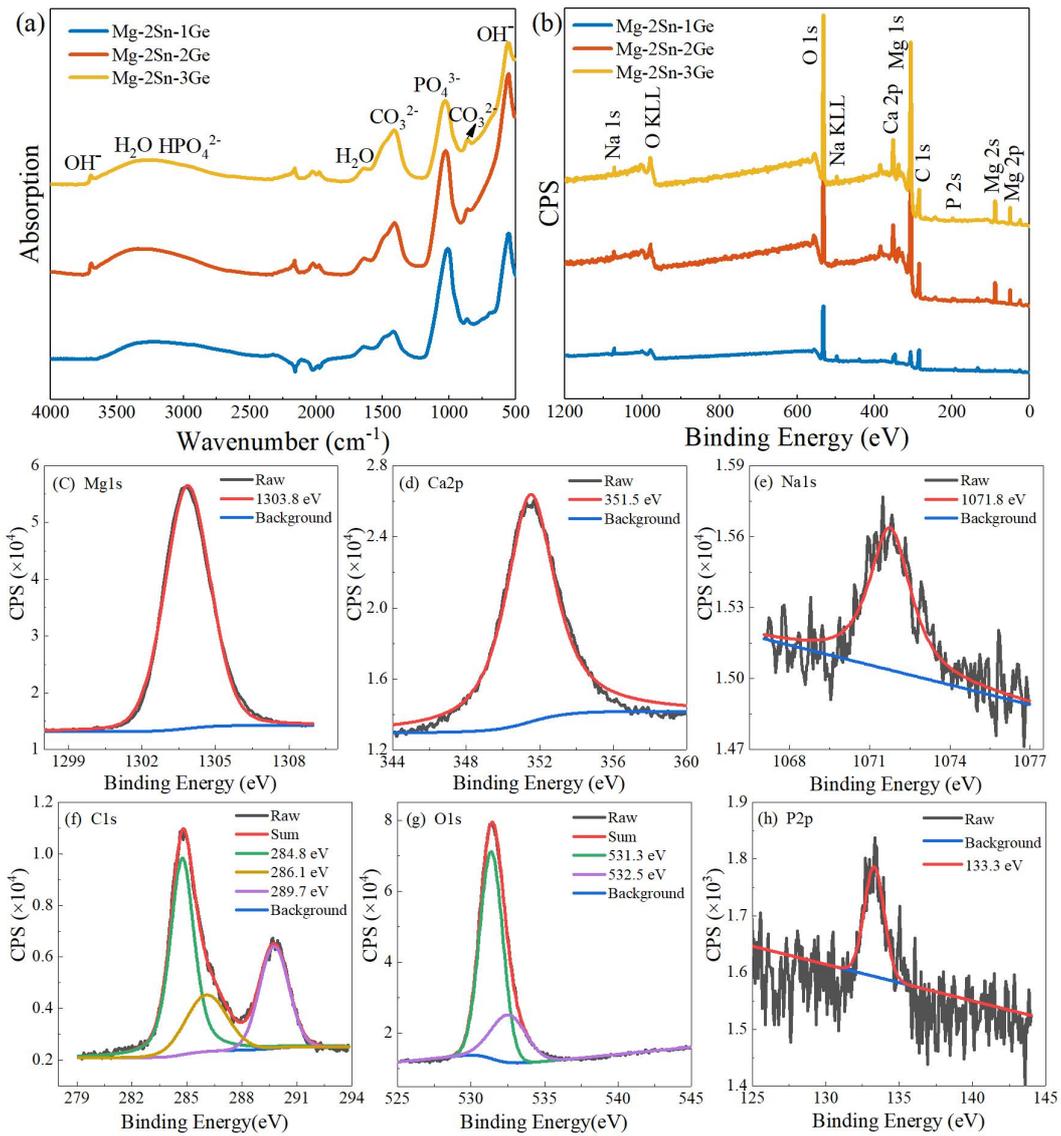

Fig. 9 (a) FTIR spectra and (b) XPS analysis of the corroded surface for the Mg-2Sn-xGe alloys after soaking for 20 days; (c-h) high-resolution XPS spectra of Mg 1s, Ca 2p, Na 1s, C 1s, O 1s, and P 2p for the Mg-2Sn-1Ge alloy.

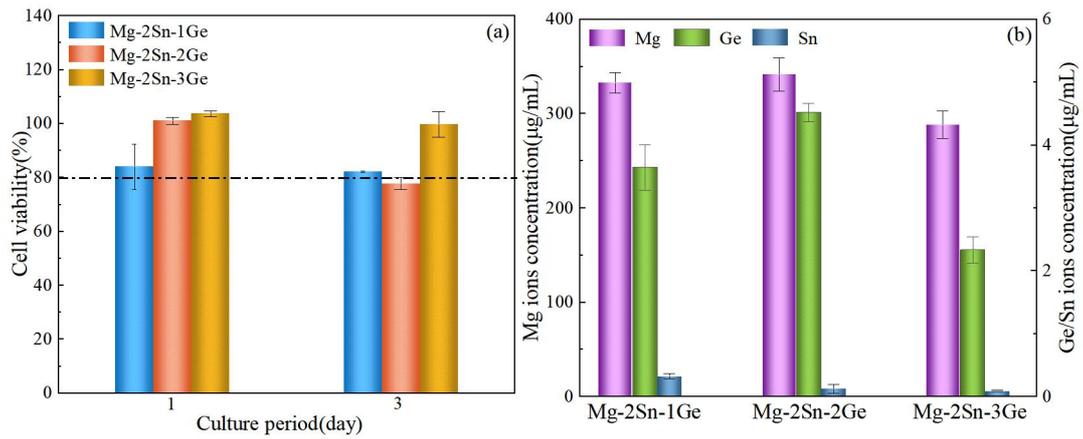

Fig. 10 (a) The viability of MC3Te3-E1 cells and (b) ions concentrations of the extracts from the Mg-2Sn-xGe alloys.

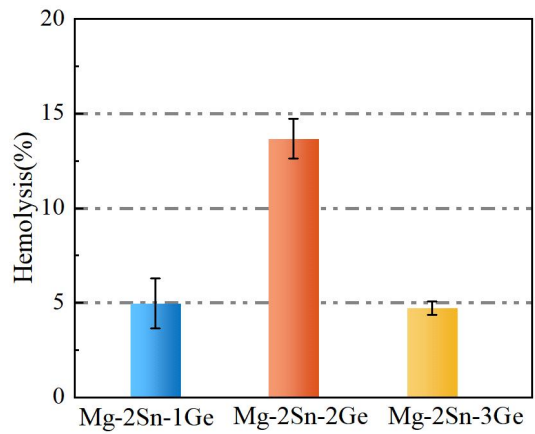

Fig. 11 Hemolysis rates of the Mg-2Sn-xGe alloys.